# Neural Network-based Power Flow Model

Thuan Pham
*Student Member, IEEE*
Department of Electrical and Computer Engineering
University of Houston
Houston, TX, USA
tdpham7@cougarnet.uh.edu

Xingpeng Li
*Member, IEEE*
Department of Electrical and Computer Engineering
University of Houston
Houston, TX, USA
xli82@uh.edu

*Abstract*—Power flow analysis is used to evaluate the flow of electricity in the power system network. Power flow calculation is used to determine the steady-state variables of the system, such as the voltage magnitude / phase angle of each bus and the active/reactive power flow on each branch. The DC power flow model is a popular linear power flow model that is widely used in the power industry. Although it is fast and robust, it may lead to inaccurate line flow results for some transmission lines. Since renewable energy sources such as solar farm or offshore wind farm are usually located far away from the main grid, accurate line flow results on these critical lines are essential for power flow analysis due to the unpredictable nature of renewable energy. Data-driven methods can be used to partially addressed these inaccuracies by taking advantage of historical grid profiles. In this paper, a neural network (NN) model is trained to predict power flow results using historical power system data. Although the training process may take time, once trained, it is very fast to estimate line flows. A comprehensive performance analysis between the proposed NN-based power flow model and the traditional DC power flow model is conducted. It can be concluded that the proposed NN-based power flow model can find solutions quickly and more accurately than DC power flow model.

## I. INTRODUCTION

For any power system, power flow is necessary to analyze the steady-state of the system. AC Power flow problems are usually solved using iterative methods such as Gauss-Seidel (GS) method and Newton-Raphson (NR). These methods can provide steady-state solutions for the system within specified accuracy boundaries. However, for a large-scale system, with hundreds of thousands of buses, it is not practical to use AC power flow equation-based methods for decision making or fast screening of the system, especially system with multiple solar or wind farms. Renewable sources provide clean, environmentally friendly energy to the power system. However, due to the inherent intermittent nature of renewable energy, novel solutions to mitigate uncertainties from renewable sources are desirable. Outside of the challenges of securing green and clean energy sources, the power industry must face difficult problems of integrating renewable energy sources to the existing aging power infrastructure. Problems will arise in terms of economic feasibility, reliability, and security. Applying ML to solve these challenging situations is critical for development of clean and green energy of the future. Applications of machine learning (ML) toward renewable energy have been widely researched and studied in recent years [1]- [2]. For example, using recursive neural networks, researchers had tried to forecast energy prices of renewable energy sources for day-ahead energy market [1]. Alternatively, using ML as an advanced algorithm to predict generation of renewable sources has been proposed [3] [4]. Deep reinforcement learning was investigated as a possible control strategy for power systems with multiple renewable energy sources [5].

Thus, it can be concluded that using machine learning as an alternative approach for solving problems involving renewable energy in power systems can be quite beneficial. Studies focusing on applications of ML show that ML algorithms could be used to gain an advantage over traditional methods in addressing various issues in power systems. Compared to traditional computational approaches, machine learning algorithms have an intrinsic generalization capability with greater computational efficiency and scalability [6]. Machine learning algorithms also have the ability to learn complex nonlinear input-output relationships, use sequential training procedures, and reflexively adapt themselves to the data.

Machine learning has been used as an algorithm selector that chooses between different power flow management algorithms depending on the state of the network [7]. Instead of directly using ML methods to solve for power flow, [7] looked at the performance benefits of using machine learning in choosing the best algorithms for power flow management. In [8], a neural network (NN) model was proposed to predict voltage magnitude and active power flow of the IEEE 14-bus system. Although it proves that NN model is an efficient method for calculating power flow, the results were only shown for a selected number of buses and branches. For [9], predictions using an NN model were made for voltage magnitude and phase angle at each bus. However, a thorough analysis regarding the performance of the NN model vs. the non-iterative DC power flow method was not conducted. Moreover, predictions for active power flow for each branch were not considered. The NN models studied in [8] and [9] were both trained and tested against only one system. This restriction does not show how well the NN model would adapt against different power systems with different bus-branch configurations. Due to the nature of machine learning algorithms to be able to reflexively adapt to different data/environments, these models did not fully explore the potential of machine learning in solving power flow.

In this paper, we developed an ML algorithm that can solve power flow solutions quickly and accurately using an NN model. Detailed research was performed regarding model selection and how to maximize the performance of the proposed NN model. The model was trained and tested against multiple systems to assess how accurate it could predict the outputs. Comparing to DC power flow model, its performance and effectiveness are evaluated and demonstrated.

The rest of the paper is organized as follows. Background analysis of power flow and machine learning are covered in section II. Section III introduces how the NN model was developed, and sample data were generated. Section IV provides analysis of the results and evaluates the performance of the proposed NN method. Section V concludes the paper and section VI describes possible future work.

## II. Preliminaries

### A. Power Flow

Power flow problem is represented by a set of nodal power balance equations that simply state that the sum of active and reactive power at a node must be equal to zeros respectively, which follows the law of conservation of energy [10]. The nodal power balance equations are listed as follows [11]

$$P_i - \sum_{k=1}^{N} |V_i| |V_k|(G_{ik} \cos \theta_{ik} + B_{ik} \sin \theta_{ik}) = 0 \quad (1)$$
$$Q_i - \sum_{k=1}^{N} |V_i| |V_k|(G_{ik} \sin \theta_{ik} - B_{ik} \cos \theta_{ik}) = 0 \quad (2)$$

where, $P_i$ and $Q_i$ are the active and reactive power injections at each node respectively. The summation terms represent the active or reactive power injections or withdrawals from the electrical network at a given node $i$. $|V_i|$ and $|V_k|$ are the voltage magnitude for the two end buses of a branch. $G_{ik}$ and $B_{ik}$ are the corresponding conductance and susceptance of a branch. The phase angle $\theta_{ik}$ is the difference in voltage phase angles of the two end buses of a branch.

Using traditional AC power flow (ACPF) methods such as NR method and GS method, an accurate steady-state solution can be found. But, due to the inherent and complex iterative nature of these algorithms, it is not beneficial for some online monitoring applications or to be integrated into optimization-based scheduling and dispatching models [12]. An alternative non-iterative method called DC power flow (DCPF) can be used in such cases. For this method, at all nodes, the voltage magnitude is assumed to be 1, and reactive power is ignored [13]. With DCPF, the solutions to steady-state active power flow can be found very quickly. However, the values may be very inaccurate due to the assumptions made earlier. Thus, it is desirable to develop a new method that can provide fast and accurate power flow solutions.

### B. Machine Learning

Machine learning is a computer algorithm that can automatically improve/learn through experience by using historical data. Models will be built using sample training data to make predictions or decisions without being explicitly programmed.

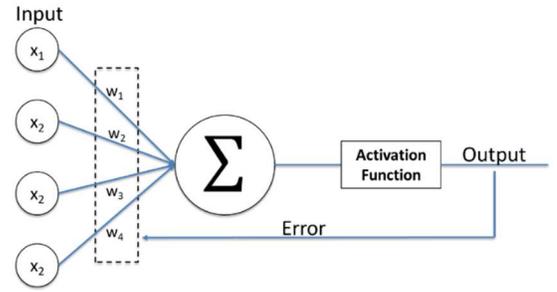

Fig. 1. Example of a simple machine learning model.

For a basic ML model in Fig. 1, training data (input) are multiplied first with a weight (*w*) vector. Then, the results are mapped to an output value after applying an activation function. Depending on model selection, different activation functions can be used. During the training process, the *w* vector is repeatedly updated until the error is less than a targeted threshold or a specified number of epochs is reached [14].

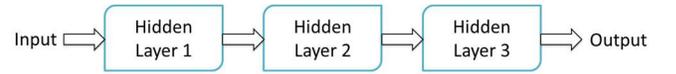

Fig. 2. Example of a multilayer neural network model.

An NN model typically involves multiple nodes (artificial neurons) with several hidden layers, such as the one from Fig. 2. The connections between nodes (vector weights) reflect the signal strength between each neuron. A complex NN with different layers performs different transformations of the input data. Each layer/node learns different features from the input data. An NN model can learn complex input-output relationships that can be difficult to comprehend or program using traditional algorithms.

The applications of ML methods are numerous across different industries, businesses, and research areas. One exciting application is to use artificial intelligence to automate the translation of text/speech. In addition, super-resolution enhancement of pictures is an interesting subject for machine learning [15]. As applications for machine learning are increasingly being explored and utilized, continuous development of machine learning is critical to solving existing and emerging challenging problems in power systems.

## III. Model Selection

For the power flow problem studied in this paper, supervised learning through regression analysis was selected as the ML method to train the NN model with a large amount of data. The matpower module for MATLAB was used to solve AC power flow problems under various power system conditions and data samples (power flow results) were collected. Initial voltage and demand were varied within ±10% of the base values. For each test power system, at least 10,000 samples were generated. The data set was divided into three groups: 80% for training, 10% for validation, and 10% for testing.

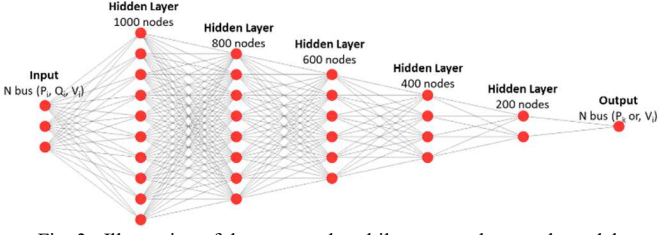

Fig. 3. Illustration of the proposed multilayer neural network model.

The proposed NN model as shown in Fig. 3 has five hidden layers. The Leaky Rectified Linear Unit (Leaky ReLU), as shown in Fig. 4, was chosen as the activation function for backpropagation. It allows a small, non-zero constant gradient to pass through to minimize the vanishing gradient problem during the training phase of a NN model.

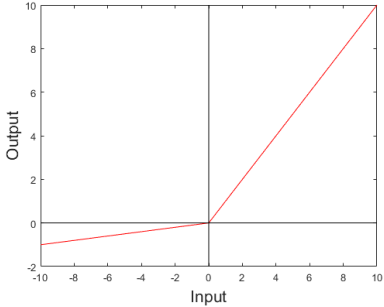

Fig. 4. Plot of leaky RELU function.

Input data for the proposed NN model include the initial voltage magnitude, the sum of active power injection, and reactive power injection for each bus. The output data for training will be steady-state voltage magnitude for each bus and active power flow for each branch.

Before feeding input data into the model for training, the data are normalized using the min-max normalization method. For every data feature, the minimum value is 0 and the maximum value is 1.

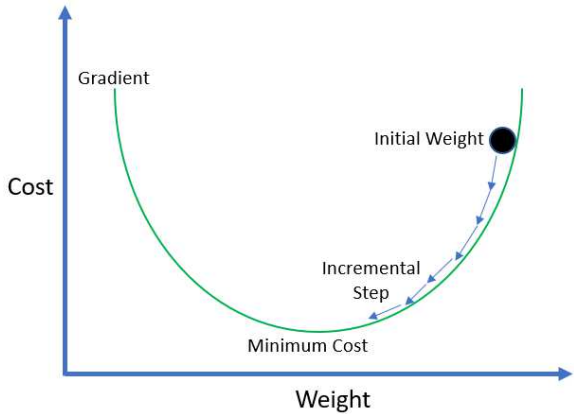

Fig. 5. Illustration of stochastic gradient descent (SGD) algorithm

Stochastic gradient descent (SGD) was chosen as the optimizing function for the NN model. SGD is very common for ML algorithm. In Fig. 5, a random starting point was chosen on the gradient curve of the objective cost function. At each subsequent iteration, the gradient was updated until the global minimum/threshold has been reached. Regular stochastic gradient requires computation of all data points for each iteration. For SGD, the selection of random samples in each minibatch for training can minimize computing time compared to using a complete training data set. Thus, it is desirable to use SGD as an optimization function.

Mean square error (MSE) loss was used to measure the performance of the model during the training process. MSE loss is commonly used for regression training of machine learning. The MSE loss function used in this paper is defined in (3), where $y_i$ denotes the actual output value while $\tilde{y}_i$ denotes the NN model estimated output value, and $n$ denotes the number of sample points.

$$MSE = \frac{1}{n}\sum_{i=1}^{n}(y_i - \tilde{y}_i)^2 \quad (3)$$

## IV. RESULTS ANALYSIS

The proposed NN model was trained to predict voltage magnitude and active power flow on multiple systems with different sizes: the IEEE 9-bus test system, IEEE 24-bus test system, IEEE 39-bus test system, IEEE 57-bus test system, and the IEEE 118-bus test system.

For comparison between the proposed NN model and DCPF model, absolute value difference and percent relative difference are used as evaluation metrics. Absolute value difference is defined as the Mean Absolute Error (MAE) between the target value against a reference value. The target values for comparison are the NN model's predictions and DCPF results. The reference base value is obtained from the full ACPF model.

$$MAE = \frac{\sum_{i=1}^{n}|x_i - y_i|}{n} \quad (4)$$

where $x$ is the target value and $y$ is the reference value.

Percent relative difference (PRD) was calculated using the relative difference formula shown below:

$$PRD = \frac{\sum_{i=1}^{n}\frac{2|x_i-y_i|}{|x_i+y_i|}}{n} * 100\% \quad (5)$$

This PRD metric is a dimensionless unit. The penalty of having a much smaller reference value relative to the target value is minimized when dividing by the sum of both values.

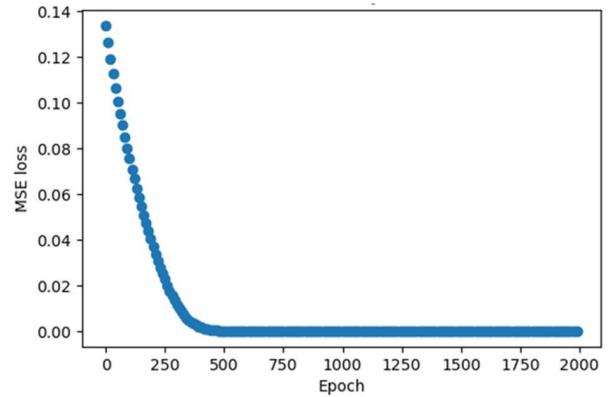

Fig. 6. MSE loss during training for active power flow of a 24-bus system.

In Fig. 6, the proposed NN model performs well in minimizing the MSE loss. The training loss decreases rapidly and then reaches the plateau zone after around 400 epochs. In addition, the model did an excellent job in fitting the predicted value toward the output value. The validation error rate curve nearly matched the training error rate curve in Fig. 7. The trained NN model does not seem to suffer from being underfitted or overfitted.

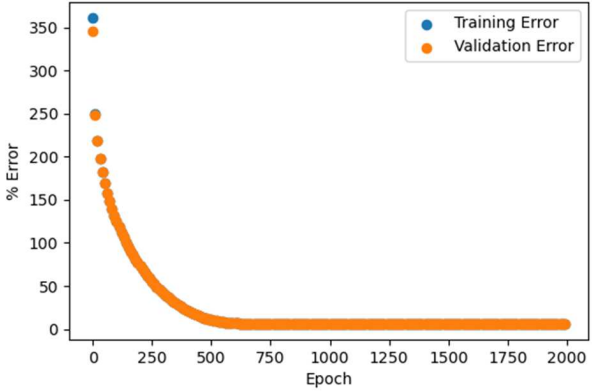

Fig. 7. MAE during training for active power flow of a 24-bus system

The proposed NN model predicts voltage magnitude at each bus for multiple systems with high accuracy: the error is around 0.5% as shown in Table I. However, since nearly all voltage magnitude values fall between a narrow range from 0.9 p.u. to 1.1 p.u., it was considerably easier for the NN model to learn and predict the output values.

TABLE I
Mean PRD of voltage magnitude of the proposed NN model on different test systems.

| Voltage Magnitude | |
|---|---|
| System | Percent Error |
| 9 Bus | 0.54 |
| 24 Bus | 0.60 |
| 39 Bus | 0.50 |
| 57 Bus | 0.33 |
| 118 Bus | 0.59 |

Table II shows the absolute errors (in MW) for line flows estimated with the proposed NN model and the traditional DCPF model. The active line power flow predicted by the proposed NN model is much more accurate than the traditional DCPF model. The lines flows (MW) predicted with the proposed NN model are much closer to the reference values from the full exact ACPF model across all measured statistical values.

TABLE II
Absolute value (MW) difference of active power flow comparing the proposed NN model and the traditional DCPF model on different test systems.

| System | Model | Mean | Max | Min | Median | Std.Dev. |
|---|---|---|---|---|---|---|
| 9 Bus | **NN** | **1.70** | **6.54** | **0.00** | **1.50** | **1.22** |
| | DCPF | 2.96 | 14.52 | 0.00 | 2.53 | 2.21 |
| 24 Bus | **NN** | **2.77** | **21.60** | **0.00** | **2.26** | **2.27** |
| | DCPF | 5.92 | 43.38 | 0.00 | 4.29 | 5.53 |
| 39 Bus | **NN** | **7.07** | **45.69** | **0.00** | **5.27** | **6.65** |
| | DCPF | 13.33 | 100.86 | 0.00 | 9.93 | 12.20 |
| 57 Bus | **NN** | **0.54** | **11.08** | **0.00** | **0.27** | **0.73** |
| | DCPF | 1.65 | 23.19 | 0.00 | 0.96 | 1.98 |
| 118 Bus | **NN** | **1.16** | **20.86** | **0.00** | **0.84** | **1.27** |
| | DCPF | 3.43 | 70.62 | 0.00 | 2.02 | 5.27 |

Table III presents the results associated with only highly loaded branches using different threshold levels. The DCPF results are not as accurate as the proposed NN model. For example, for all branches with actual active power flow over 200 MW for the IEEE 24-bus system, the PRD mean for the proposed NN model is only 0.65%, which is very close to the actual results reported from the full ACPF model. However, the PRD is 1.35% for the DCPF model. It is double the error of the proposed NN model.

TABLE III
PRD of active power flow comparing the proposed NN model and the DCPF model for different threshold levels on the IEEE 24-bus system.

| System | Model | Mean | Max | Min | Median | Std.Dev. |
|---|---|---|---|---|---|---|
| 50 MW | **NN** | **1.72** | **17.14** | **0.00** | **1.26** | **1.62** |
| | DCPF | 6.42 | 75.28 | 0.00 | 3.43 | 10.02 |
| 100 MW | **NN** | **1.42** | **8.95** | **0.00** | **0.89** | **1.52** |
| | DCPF | 6.55 | 75.28 | 0.00 | 2.75 | 12.76 |
| 150 MW | **NN** | **1.33** | **8.95** | **0.00** | **0.76** | **1.59** |
| | DCPF | 2.96 | 24.63 | 0.00 | 1.63 | 3.53 |
| 200 MW | **NN** | **0.65** | **3.34** | **0.00** | **0.53** | **0.52** |
| | DCPF | 1.35 | 5.74 | 0.00 | 1.22 | 0.92 |

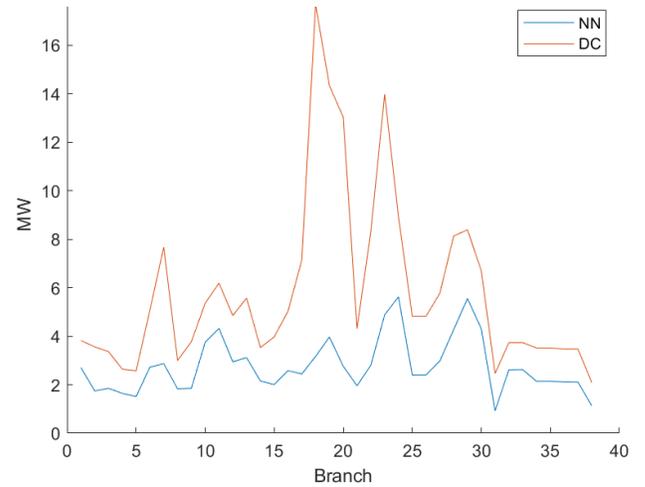

Fig. 8. Mean absolute value difference for active power flow of a 24-bus system.

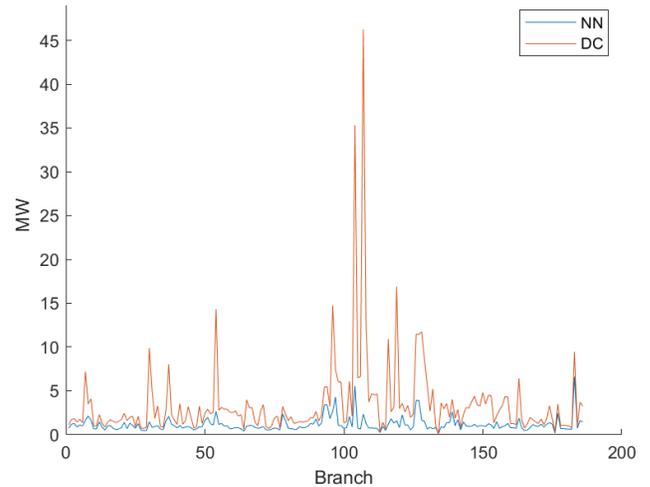

Fig. 9. Mean absolute value difference for active power flow of a 118-bus system.

Fig. 8 and Fig. 9 present the mean absolute error of line power flows on the IEEE 24-bus system and IEEE 118-bus system respectively. From these two figures, it is observed that the proposed NN model outperforms the traditional DCPF model by a large margin. Note that the DCPF model has large outliers with great errors while the proposed NN model does not. For the proposed NN model, the predicted line active power flows are rarely more than 5MW away from the true values. It can be concluded that the proposed NN model is superior to the traditional DCPF model in terms of accuracy.

## V. Conclusion

Based on the result, it can be concluded that the proposed NN model produces much better results at predicting line active power flows compared to the traditional DCPF model. In addition, the voltage magnitudes predicted by the proposed NN model closely match the results of ACPF. The potential of using the NN model as a supervised machine learning algorithm to calculate power flow has been explored in this paper. Unlike DCPF, no assumption must be made regarding voltage magnitudes and the proposed NN model can predict the values of voltage magnitudes very accurately. The values of line active power flows using the proposed NN model are very close to solutions solved by ACPF. In addition, solutions to both line active power flows and voltage magnitudes can be found quickly once the model has been trained offline. The proposed NN model performs faster with better results compared to the DCPF model.

Model selection is an important step for the ML algorithm. In addition, training the NN model may require a large data set of samples and computing resources. Building and training NN models can be difficult and computationally expensive, but the reward is significant. Applications of NN in power systems are still being explored and researched [1], [5]. Regardless, the potentials of using ML to build and improve upon previous power systems technologies are limitless.

## VI. Future Work

The proposed NN model shows potentially useful applications of machine learning in power systems. Due to the adaptability nature of ML, NN predictions can be used as part of the optimization model to solve optimal power flow (OPF) for system with multiple renewable sources. With the NN model, we can directly calculate line flows as part of the constraint and simplify OPF calculation. Alternatively, we can also develop a different machine learning algorithm to account for unpredictability of renewable energy in OPF calculation. Another potential application of the proposed NN model is to quickly screen a long list of contingencies and filter out non-critical contingencies for large-scale power systems in real-time.

## VII. References

[1] P. Mandal, T. Senjyu, N. Urasaki, A. Yona, T. Funabashi and A. K. Srivastava, "Price Forecasting for Day-Ahead Electricity Market Using Recursive Neural Network," in *IEEE Power Engineering Society General Meeting*, 2007.

[2] "Long-Term Recurrent Convolutional Network-based Inertia Estimation using Ambient Measurements," *IEEE PES General Meeting (under review),* 2022.

[3] L. Wu, J. Park, J. Choi, J. Cha and K. Y. Lee, "A study on wind speed prediction using artificial neural network at Jeju Island in Korea," in *Transmission & Distribution Conference & Exposition: Asia and Pacific*, Seoul, Korea, 2009.

[4] "Solar Power Output Prediction Using Multilayered Feedforward Neural Network: A Case Study of Jaipur," in *IEEE International Symposium on Sustainable Energy, Signal Processing and Cyber Security* , Gunupur Odisha, India , 2020.

[5] S. Huang, M. Yang, C. Zhang, J. Yun, Y. Gao and P. Li, "A Control Strategy Based on Deep Reinforcement Learning Under the Combined Wind-Solar Storage System," in *IEEE 3rd Student Conference on Electrical Machines and Systems*, 2020.

[6] S. M. Miraftabzadeh, F. Foiadelli, M. Longo and M. Pasetti, "A Survey of Machine Learning Applications for Power System Analytics," in *International Conference on Environment and Electrical Engineering*, 2019.

[7] J. E. King, S. C. E. Jupe and P. C. Taylor, "Network State-Based Algorithm Selection for Power Flow Management Using Machine Learning," *IEEE Transactions on Power Systems,* vol. 30, no. 5, Sept. 2015.

[8] M. Zhou, "Exploring Application of Machine Learning to Power System Analysis," ResearchGate, 218.

[9] W. A. Alsulami and R. S. Kumar, "Artificial Neural Network based Load Flow Solution of Saudi National Grid," in *Saudi Arabia Smart Grid Conference*, 2017.

[10] M. S. S. T. O. A. B. J. Duncan Glover, Power System Analysis and Design, Cengage Learning, 2022.

[11] A. S. K. P. B. Xingpeng Li, "Sensitivity factors based transmission network topology control for violation relief," *IET Generation, Transmission & Distribution,* vol. 14, no. 17, pp. 3539-3547, July 2020.

[12] A. Keyhani, A. Abur and S. Hao, "Evaluation of Power Flow Techniques for Personal Computers," in *IEEE Transactions on Power Systems*, 1989.

[13] Y. Qi, D. Shi and D. Tylavsky, "Impact of assumptions on DC power flow model accuracy," in *North American Power Symposium*, 2012.

[14] R. Lukomski and K. Wilkosz, "Power System Topology Verification Using Artificial Neural Network Utilization of Measurement Data," in *IEEE Bologna Power Tech Conference Proceedings*, 2003.

[15] J. H. W. C. T. S. D. J. F. M. N. Chitwan Saharia, "Image Super-Resolution via Iterative Refinement," in *International Conference on Computer Vision*, 2021.

[16] A. V. Ramesh and X. Li, "Machine Learning Assisted Approach for Security-Constrained Unit Commitment," *IEEE Transactions on Power Systems (under review)*.